\documentclass[twocolumn]{aastex63}
\usepackage[utf8]{inputenc}
\usepackage{lineno}

\usepackage{mathrsfs}
\usepackage{amsmath}
\usepackage{booktabs}
\ifdefined\DeclareUnicodeCharacter
  \DeclareUnicodeCharacter{FB01}{\nobreakspace}
\fi
\ifdefined\DeclareUnicodeCharacter
  \DeclareUnicodeCharacter{FB02}{\nobreakspace}
\fi
\newcommand{\jcafont}{
  \color{red}
}
\DeclareTextFontCommand{\jca}{\jcafont}
\usepackage[normalem]{ulem}
\newcommand\jcaout{\bgroup\markoverwith
{\textcolor{red}{\rule[.5ex]{2pt}{0.4pt}}}\ULon}

\newcommand{\mafont}{
  \color{blue}
}
\DeclareTextFontCommand{\ma}{\mafont}
\usepackage[normalem]{ulem}
\newcommand\maout{\bgroup\markoverwith
{\textcolor{blue}{\rule[.5ex]{2pt}{0.4pt}}}\ULon}

\received{\today}
\revised{}
\accepted{}

\shorttitle{Runaway return current}
\shortauthors{Alaoui et al.}

\graphicspath{{./}{figures/}}

\begin{document}
%\linenumbers
\title{Role of suprathermal runaway electrons returning to the acceleration region in solar flares}
\correspondingauthor{Meriem Alaoui}
\email{meriem.alaouiabdallaoui@nasa.gov}

\author[0000-0003-2932-3623]{Meriem Alaoui}
\affiliation{The Catholic University of America\\
620 Michigan Ave, Washington, DC 20064, USA}
\affiliation{NASA Goddard Space Flight Center\\8800 Greenbelt Rd, Greenbelt, MD 20771}

\author{Gordon D. Holman}
\affiliation{NASA Goddard Space Flight Center\\8800 Greenbelt Rd, Greenbelt, MD 20771}

\author{Joel C. Allred}
\affiliation{NASA Goddard Space Flight Center\\8800 Greenbelt Rd, Greenbelt, MD 20771}

\author{Rafael T. Eufrasio}
\affiliation{University of Arkansas, Fayetteville, AR 72701}

\begin{abstract}
During solar flares, a large flux of energetic electrons propagate from the tops of reconnecting magnetic flux tubes toward the lower atmosphere. Over the course of the electrons' transport, a co-spatial counter-streaming return current is induced, thereby balancing the current density. In response to the return current electric field, a fraction of the ambient electrons will be accelerated into the runaway regime. However, models describing the accelerated electron beam/return-current system have generally failed to take these suprathermal runaway electrons into account self-consistently. We develop a model in which an accelerated electron beam drives a steady-state, sub-Dreicer co-spatial return-current electric field, which locally balances the direct beam current and freely accelerates a fraction of background (return-current) electrons. The model is self-consistent, i.e., the electric field induced by the co-evolution of the direct beam and the runaway current is considered. We find that (1) the return current electric field can return a significant number of suprathermal electrons to the acceleration region, where they can be further accelerated to higher energies, runaway electrons can be a few tens of percent of the return current flux returning to the nonthermal beam's acceleration region, (2) the energy gain of the suprathermal electrons can be up to $10-35$\,keV, (3) the heating rate in the corona can be reduced by an order of magnitude in comparison to models which neglect the runaway component. The results depend on the injected beam flux density, the temperature and density of the background plasma.
\end{abstract}
\keywords{Sun: flares — Sun: X-rays, gamma rays-  Runaway electrons}

\section{Introduction} \label{sec:intro}
Co-spatial return currents (RCs) have been proposed to balance the electron flux required to explain the observed X-ray bremsstrahlung emission in solar flares \citep{1976SoPh...48..197H}. The bremsstrahlung emission above $\sim$ 20 keV provides the most direct link to the nonthermal electron distribution. The RC locally neutralizes the charge build-up and cancels the magnetic field induced by the beam of accelerated electrons, thereby solving the so-called “number problem” \citep{2008LRSP....5....1B}, and the associated current stability problem \citep{1934PhRv...45..890B,1955PhRv...98.1584B}. The electric field that drives the RC  produces a potential drop between the acceleration region in the corona and the flare-loop footpoints, and decelerates the electrons in the beam. There is growing observational evidence that the acceleration region is located in the outflow jets away from the reconnection region \citep[e.g.,][]{2003ApJ...596L.251S,2018ApJ...863...83G,2018ApJ...866...62C,2021ApJ...908L..55C}. The potential drop can be detected by a flattening in the X-ray spectrum at low energies, in the deka-keV range \citep[][]{2012ApJ...745...52H,2006ApJ...651..553Z}. The electric field accelerates the thermal background electrons that then undergo collisions to produce a drifting Maxwellian \citep{1977ApJ...218..306K,1984ApJ...280..448S,1980ApJ...235.1055E,1989SoPh..120..343L,1995A&A...304..284Z,2005A&A...432.1033Z,2006ApJ...651..553Z,2007ApJ...666.1268A,2013ApJ...773..121C}. However, the electrons for which the electrical force exceeds the collisional friction force, are continuously accelerated into the runaway regime. The effect of suprathermal runaway electrons has not been taken into account in describing the dynamics of the direct beam/co-spatial return-current system. A model which describes these dynamics is developed in the weak-field (sub-Dreicer) regime in this paper. 

Runaway electrons are observed in various astrophysical and laboratory plasmas. Notable examples are thunderstorms \citep{2012SSRv..173..133D,2009PhyU...52..735G} and tokamaks \citep[][and references therein]{2019NucFu..59h3001B}. In solar flare theory, runaway electrons have been studied in the context of a parallel electric field as the accelerating mechanism of beam electrons \citep[e.g.,][]{1985ApJ...293..584H,1992ApJ...400L..79H,1994ApJ...435..469B,1995ApJ...452..451H}. This is different from the RC electric field since that is generated as a consequence of the propagation of the already accelerated electron beam. The RC electric field simultaneously both decelerates the direct nonthermal beam and accelerates a fraction of electrons in the background plasma. In a Fokker-Planck model of the nonthermal electron propagation in the solar atmosphere, \citet{2006ApJ...651..553Z} recognized that the return current electric field can be super-Dreicer in solar flare conditions (Figure 1 panels b and c in their paper), but did not take into account the dynamics introduced by the presence of runaway electrons on the beam/RC system. Their model and that of \citet{2020ApJ...902...16A} can however be used to calculate the fraction of nonthermal electrons in the direct beam which are scattered backward and become part of the return current. Both the suprathermal runaways and back-scattered nonthermal electrons should be studied self-consistently. This will be the subject of an upcoming paper. 

Collisionless effects including runaway electrons and current-driven instabilities of the beam/RC system were reviewed by \citet{1985A&A...142..219R}. They describe the following three limiting regimes for the propagation of the beam: (1) The RC losses are negligible when the beam flux is sufficiently low, (2) the RC losses are not negligible but the return current is stable (the regime discussed in this paper), and (3) the RC becomes unstable to current-driven instabilities \citep{2002ASSL..279.....B,2008A&A...487..337B,2017ApJ...851...78A}. 

The equations derived in this paper are restricted to sub-Dreicer electric fields  \citep{1959PhRv..115..238D}. This constraint is introduced for consistency with existing analytic results for classical collisions. We will show that under the assumptions of our model, if the density of the accelerated beam is less than 0.3 that of the background plasma, the RC electric field is sub-Dreicer everywhere along the flare loop. In addition, our current X-ray observations are limited by a 4 s time resolution for a statistically significant spectrum, which is longer than the time to reach the steady-state for most coronal conditions. It will be discussed in sections~\ref{sec:runaway_rate} and ~\ref{sec:method}.

The purpose of this paper is to investigate the contribution of suprathermal runaway electrons to the dynamics of the direct beam/RC system. In particular, we investigate the fraction of the return current that is carried by runaway electrons under different conditions, as well as  how these electrons alter the dynamics of the direct beam/return-current system. This will allow us to address more fundamental questions including:
\begin{enumerate} 
\item How are the acceleration and propagation regions coupled? Runaway electrons provide suprathermal seed particles to the acceleration region, where they can be further accelerated to higher energies.  Thus, understanding the conditions favorable for their generation is important for understanding the electron acceleration process itself.
\item How much energy is deposited and at what depths in the solar atmosphere? The RC can be used as a diagnostics tool to constrain the relative importance of beam propagation mechanisms and energy deposition.  

\item To what extent is the shape of X-ray spectra, particularly the flattening often seen at lower energies \citep[e.g.][]{2019SoPh..294..105A}, a consequence of the acceleration mechanism or of the propagation processes?
Is the spectrum of the accelerated electrons a single power-law above a low-energy cutoff or are there significant deviations from this simple assumption?
\end{enumerate}

In Section~\ref{sec:model}, we derive the equations describing the weak-field return current model. We look at the spatial evolution of the return-current electric field, the beam and return current densities, the volumetric heating rates, and the potential drop from the acceleration region to the thick target. We will build the model in three steps. First neglecting Coulomb collisions between the beam and background electrons, with a low potential drop, then an arbitrary potential drop, and finally including Coulomb collisions self-consistently. In Section~\ref{sec:application}, we compare our runaway model with solar flare models that exclude runaways to demonstrate when their inclusion becomes critical. A comparison of beam propagation in various solar atmospheres is presented in Section~\ref{sec:var_atm}. The effect of pitch-angle and return-current driven instabilities  are discussed in Section~\ref{sec:discussion}. The main results are summarized in Section~\ref{sec:summary}.

\section{Co-Spatial Return-Current Runaway Model--Weak Field Regime}
\label{sec:model}
\subsection{Runaway rate}
\label{sec:runaway_rate}

\noindent When a sub-Dreicer electric field ($\mathscr{E} < \mathscr{E}_{D}$) is applied to a thermal plasma, a fraction of the Maxwellian distribution will run away, and the bulk of the Maxwellian distribution will drift under the influence of the electric field. A higher electric field results in a higher runaway rate.
Many authors carried out solutions to the spatially homogeneous Fokker-Planck equation where an electric field is applied to a Maxwellian distribution \cite[e.g.,][for a review]{1979NucFu..19..785K} and calculated the runaway growth rate. 

In this section we benchmark the numerical growth rate from \cite{2014CoPhC.185..847L} and \cite{1973PhRvL..31..690K} with the analytical expression of \cite{1964PhFl....7..407K}. We must first adjust these growth rates to consistent definitions of the Dreicer field, collision frequency and thermal velocity.

\cite{1964PhFl....7..407K} provided an asymptotic solution to the spatially homogeneous Fokker-Planck equation by expanding the distribution function in powers of $\frac{\mathscr{E}}{\mathscr{E}_D}$ near $\mu=\frac{v_{\parallel}}{v}= 1$ by recognizing that if the electric field is much less than the Dreicer field then the critical velocity $v_{cr}$ above which electrons are in the runaway regime have a pitch-angle $\mu\sim 1$. Under these conditions, the critical velocity above which runaway occurs is given by \citep[e.g.,][]{2008PhPl...15g2502S} \begin{equation}v_{cr}=v_e \sqrt{\mathscr{E}_D/{2\mathscr{E}}}\label{eq:crit_vel},\end{equation}
where $v_e=\sqrt{2k_B T/m_e}$ is the thermal velocity.\\
\noindent The Kruskal $\&$ Bernstein runaway rate, in units s$^{-1}$, is given in \citep[e.g.,][]{1985ApJ...293..584H,1964PhFl....7..407K}, 
 
\begin{equation}
\gamma_{run}= C \nu_e  \Big( \frac{\mathscr{E}_D}{\mathscr{E}} \Big )^{3/8}  \exp \Big( -\sqrt{2 \frac{\mathscr{E}_D}{\mathscr{E}}} - \frac{1}{4} \frac{\mathscr{E}_D}{\mathscr{E}}\Big)
\label{eq:kb64}
\end{equation}
\noindent where C is a constant of order 1 \citep[e.g.,][]{2019JPlPh..85f0601H,1975NucFu..15..415C}, and

\begin{align}
\label{eq:dreicer_field}
  \mathscr{E}_D & = 4\pi e^3  \ln\Lambda \frac{n_e}{k_B T} = \frac{3\sqrt{\pi}}{2} \frac{m v_e \nu_e}{e}\nonumber \\
                & = 1.0\times 10^{-11} \ln\Lambda \frac{n_e[cm^{-3}]}{T [K]}
\end{align}
is the Dreicer field in StatV cm$^{-1}$ \cite[e.g.,][]{2012ASSL..391.....S}, 
\begin{equation}
  \nu_e=\frac{4}{3}\frac{\sqrt{2\pi}}{\sqrt{m_e}} \frac{n e^4 \ln\Lambda}{(k T_e)^{3/2}}
  \label{eq:freq}
\end{equation} 
is the electron collision frequency given by \cite{NRL...plasma..formulary}. In these expressions, $m_e$ is the electron mass, $e$ is the absolute value of the electron charge, $\ln \Lambda$ is the Coulomb logarithm, $n_e$ is the electron density, $k_B$ is the Boltzmann constant, and $T_e$ is the electron temperature.
\begin{figure}[h]
\centering
 \includegraphics[width=0.47\textwidth]{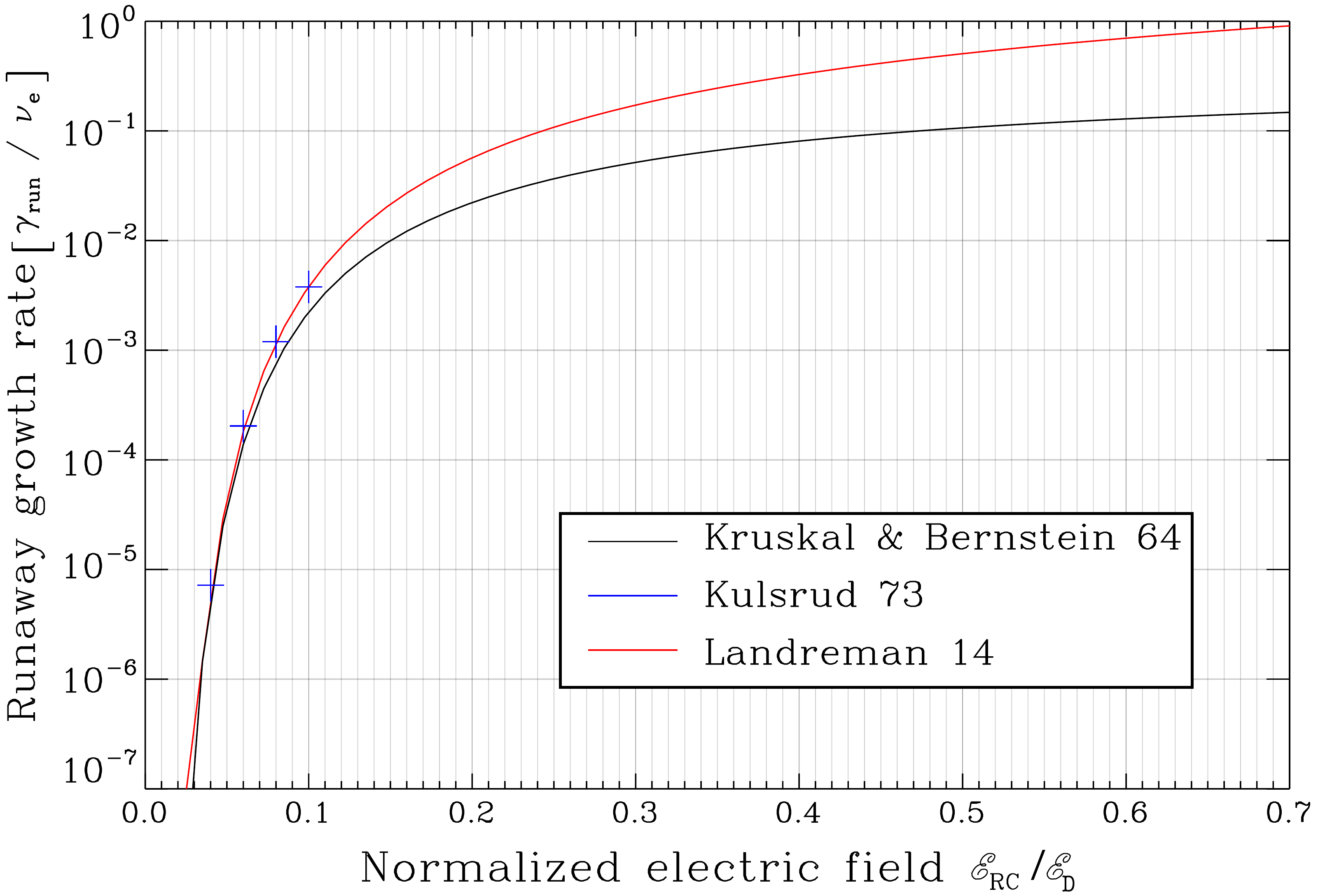}
   \caption{Normalized dimensionless runaway rate ${\gamma_{run}}/{\nu_e}= ({\nu_e n_e})^{-1}\frac{d n_r}{dt}$ from \cite{2014CoPhC.185..847L,1964PhFl....7..407K,1973PhRvL..31..690K}. }
  \label{fig:growth_rate}
\end{figure} 

\cite{1973PhRvL..31..690K} numerically solved the Fokker-Planck equation using a collision operator that is linearized, non-relativistic, spatially homogeneous and axially symmetrical in velocity space. The solutions are represented by the blue crosses in Figure~\ref{fig:growth_rate}, and the collision frequency used is $\nu_{k73}=\frac{3\sqrt{\pi}}{\sqrt{2}}\nu_e$.
\cite{2014CoPhC.185..847L} used the relativistic test-particle collision operator in the limit of non-relativistic plasma temperature \cite{2011NucFu..51d3004P}. The collision operator is valid for arbitrary electron velocities. Their operator neglects self-collisions between runaway electrons, i.e., runaways only collide with the Maxwellian bulk. The code can handle any initial distribution, not only small deviations from a Maxwellian. 

The runaway growth rate is given by Equation 17 in \cite{2014CoPhC.185..847L}.
Their results have been extended to higher fractions of the electric field from 0.1 to 0.7 in Figure~\ref{fig:growth_rate} using CODE (COllisional Distributions of Electrons) \citep[][]{2014CoPhC.185..847L,2015PhRvL.114k5002S,2018JPlPh..84a9002E}.

To sum up, the KB64 solution is equivalent to Landreman's at lower fractions of the electric field: $\mathscr{E}_{RC}/\mathscr{E}_D\lesssim 0.1$, if the constant C is equal to 1 instead of 0.35 \citep[e.g.,][]{1985ApJ...293..584H}, and the Dreicer field and collision frequency are defined by Equations~\ref{eq:dreicer_field} and ~\ref{eq:freq}, respectively.

\subsection{Assumptions}
\label{sec:method}

\noindent In this section we derive a set of equations for describing the beam/RC system in which a power-law distribution of nonthermal electrons is injected into a thermal, fully ionized hydrogen plasma modeled by a 1D loop. The assumptions used made by our model are described in the following:   

\begin{itemize}
    \item The accelerated electrons are continuously injected at the apex of a 1D loop
model. Their flux density distribution is represented by a single power-law with a sharp low-energy cutoff E$_{c0}$. That is the flux density distribution, $F$, is given by
\begin{equation}
F(E,x=0)=(\delta -1)F_{e0} E_{c0}^{\delta-1}E^{-\delta},
\end{equation} where $\delta$ is the electron spectral index, $F_{e0}$ the total injected electron flux density, and $E$ is the electron energy, and $x$ is the geometric distance from the loop apex.

    \item Steady-state:  We consider time scales much longer than that determined from the electron-ion collision frequency meaning that return current/beam system has reached a steady-state \citep{1990A&A...234..496V}. This is given by the inverse of Equation~\ref{eq:freq}. This is much less than 4\,s for usual values of coronal temperature and density combinations in solar flares. For example, for a 30 MK and 10$^{9}$\,cm$^{-3}$ plasma the collision time is 2.1 s, and 15 ms for 1 MK and 10$^9$ cm$^{-3}$ plasma. We are in addition limited by the 4\,s time resolution of RHESSI \citep[][]{2002SoPh..210....3L} for a statistically significant spectrum.

\item 1D loop: All electrons have velocities parallel to the magnetic field. This is useful to provide direct comparisons with previous results of \cite{2012ApJ...745...52H} and AH17, and gain physical insight into the problem. We will assume the temperature and density to be constant in the corona in Section~\ref{sec:application} and a model with temperature and density stratification in section~\ref{sec:pitch-angle}.

   \item  We focus on modeling the effect of return-current and collisional losses on the dynamics of the beam/RC system. Other mechanisms not included in this paper which affect the propagation of the nonthermal beam include magnetic mirroring \cite[][]{1993A&A...278..627K,1995A&A...304..284Z}, and wave-particle instabilities \cite[e.g.,][]{2009ApJ...707L..45H,2011A&A...529A.109H}.
    
    \item The plasma resistivity, $\eta$, is assumed to be the classical Spitzer resistivity, $\eta_S$ \citep{1962pfig.book.....S,1953PhRv...89..977S}.

    \item We assume that beam electrons are thermalized and lost from the beam when they have energy less than $E_{th} = \delta kT$, where T is the ambient plasma temperature \citep{2015ApJ...809...35K,2019ApJ...880..136J}. \cite{2020ApJ...902...16A} showed that energy diffusion dominates over the dynamic friction, i.e., thermalization dominates over energy loss for an electron energy below $(\delta+1) kT$.

     \item We assume that there is not a neutralizing ion beam propagating co-spatially along with the electron beam. Since ions are three orders of magnitude more massive, their velocity and, therefore, current density is likely much smaller than the electron beam's. Therefore, even if considered, their neutralizing effect on the electron beam would be small  \citep[see reviews by][]{1997JGR...10214631M,2016JGRA..12111667H}.

\end{itemize}

\subsection{Method of analysis: Negligible collisional losses}
In a low-density corona, collisions between beam electrons and the background plasma are relatively infrequent, but the return current electric field can produce a potential drop that is a significant fraction of the beam electrons' energies. Therefore, we assume that in the corona the energy lost by beam electrons due to the return current electric field dominates over energy lost from Coulomb collisions.

We make the assumption that in the steady state the return current locally balances the beam current. That is 
\begin{equation}|j_{RC}(x)| = |j_{b}(x)|,\label{eq:balance}
\end{equation}
where $j_b$ is the beam current density, and $j_{RC}$ is the return current density, which can be written as the sum of a drifting bulk thermal distribution ($j_{d}$) and the runaway tail of the thermal distribution ($j_{run}$): \begin{equation}j_{RC}(x)= j_{d}(x)+ j_{run}(x)\end{equation} 

The electron beam current density, $j_b = e F_{e}$, in the presence of a return current electric field was studied by \citet{2012ApJ...745...52H}. He derived an analytic expression describing the beam flux density, $F_e(x)$, as a function of distance along a loop, given the injection of a power-law distribution of electrons. Here we reproduce that result (his Equation 15) for clarity,
 \begin{equation}
 F_{e}(x, E_{c0})= \begin{cases}
F_{e0} & \mbox {; } x \leq x_{RC} \\
F_{e0}{\Big[ { \frac{{E_{th}+ e\int_0^x \mathscr{E}_{RC}(x') dx'}}  {E_{c0} }}  \Big]^{1-\delta}}   & \mbox{; } x >x_{RC}  \end{cases} \label{eq:beamflux} \end{equation} 
 where $\delta$ is the electron distribution spectral index, $F_{e0}$ is the injected electron flux density in electrons cm$^{-2}$ s$^{-1}$, $E_{c0}$ is the low-energy cutoff of the injected electron distribution, $x_{RC}$ is the distance along the loop where beam electrons are thermalized. Since electrons with an initial energy equal to the low-energy cutoff are thermalized first, $x_{RC}$ is defined as the distance where the potential drop becomes equal to  ${E_{c0} -E_{th}}$, i.e.,  $ e\int_0^{x_{RC}} \mathscr{E}_{RC} (x) dx= E_{c0}-E_{th}$.

The number density of runaway electrons produced per second at each position $x$ is $n_e(x)\gamma_{run}(x)$.  These electrons are all accelerated upward by the return-current electric field.  All runaways produced below position $x$ contribute to the runaway current at $x$.  For our assumed steady-state solution and constant plasma density, the runaway flux and, therefore, current is:
\begin{equation}
  j_{run}(x)=e n_e \int_L^x \gamma_{run}(x') dx'
  \label{eq:jrunaway}
\end{equation} 
where $\frac{\gamma_{run}}{\nu_e}=\frac{1}{\nu_e n_e}\frac{dn_{run}}{dt}$ is the numerical runaway rate of \cite{2014CoPhC.185..847L} as shown in Figure~\ref{fig:growth_rate}.

For time scales much longer than the collisional time scale, we assume the drifting component of the return current is Ohmic, as supported by \citet{1990A&A...234..496V}, and is given by
\begin{equation}
    j_d(x) = \frac{\mathscr{E}_{RC}(x)}{\eta_s}\label{eq:jdrift}
\end{equation}

\subsubsection{Small potential drop: No thermalization}\label{sec:notherm}

\noindent Let us start with the simple case where the length of the loop is less than $x_{RC}$.Substituting the expressions for $j_b$, $j_{run}$, and $j_d$ into Equation~\ref{eq:balance} gives
\begin{equation}    
\begin{split} {e F_{e0}} ={\frac{\mathscr{E}_{RC}(x)} {\eta_s}} -  e \int_x^L n_e (x') \gamma_{run} (x') dx' 
\end{split}
\label{eq:first_trans}
 \end{equation}

Multiplying by $\mathscr{E}_D/\eta_s$ and setting $z(x)=\mathscr{E}_{RC}(x)/\mathscr{E}_D(x)$, we obtain the dimensionless expression:
\begin{equation}    
 \begin{split}{\frac{ e F_{e0}\eta_s}{\mathscr{E}_D}}  = z(x) + {\frac{e \eta_s}{\mathscr{E}_D}}  \int_L^x n_e(x') \gamma_{run}(x') dx' 
 \end{split}
 \label{eq:508}
 \end{equation}
 
We solve for $z$ in Equation~\ref{eq:508} using the boundary condition at the footpoints that $j_{run} = 0$, so \begin{equation} z(x=L)= {\frac{e F_{e0}  \eta_s(L)}{\mathscr{E}_D(L)}}.\end{equation}
The return current heats the ambient plasma via Joule heating. That is the resistive dissipation of energy from the drifting component. We assume that the runaway component does not significantly heat the ambient plasma. 

\subsubsection{Large potential drop: Electron thermalization}
 \label{sec:RA.RC}
Now let us consider the case where the loop length is greater than $x_{RC}$. 
As in the previous section, substituting the expressions for $j_b$, $j_{run}$, and $j_d$ into Equation~\ref{eq:balance}, multiplying by ${\eta_s}/{\mathscr{E}_{D}}$, and using $z(x)={{\mathscr{E}_{RC}(x)} /{\mathscr{E}_{D}(x)}}$, we obtain the dimensionless equation:

\begin{equation} 
\begin{split} 
{ \frac{eF_{e0} \eta_s(x)}{\mathscr{E}_D(x)}}  \Big[{ {\frac{E_{th}(x)} {E_{c0}}} + {\frac{e}{E_{c0}}} \int_0^x {\mathscr{E}_D(x') z(x') {dx'} }   }\Big]^{1-\delta} \\= z(x) -  \frac{e \eta_s(x)}{\mathscr{E}_D(x)}\int_x^L n_e(x') \gamma_{run}(x')  dx' 
\end{split} 
\label{eq:510}
\end{equation}
 
In this case, the boundary condition and thermalization length are not known \textit{a priori}, because the solution at any position along the loop is sensitive to the behavior of the runaway current which is integrated from the footpoints upward, whereas the beam current is integrated from the looptop downward. The dynamics of the runaways inform the dynamics of the beam, and vice-versa. 

We solve this equation iteratively. The solution without thermalization is used to determine the critical values of the low-energy cutoff $E_{cc}$ and injected electron flux density $F_{e0c}$ for which the thermalization distance is $x_{rc}=L$. For all values of the low-energy cutoff $E_{c0} < E_{cc}$ and $F_{e0}> F_{e0c}$, Equation~\ref{eq:510} should be used. 
Using Equation~\ref{eq:508} as initial guess, and noticing that the drift current is proportional to $z$, this value is updated at each iteration as to minimize $|j_{b}-j_{RC}|$ everywhere along the loop until a satisfactory tolerance is achieved. As in the previous section, the return current heating rate in this case is simply Joule heating from the drifting component. For simplicity we will refer to this model as RA.RC which stands for the return-current model with runaway acceleration, the RC model without runaways presented in  \citet{2012ApJ...745...52H} is labeled H12.

\subsection{Self-consistent solution of return-current and collisional losses}
\label{sec:RA.RC.CC}
 Since energy lost by beam electrons due to Coulomb collisions (CC) is sensitive to the electrons' energies and the ambient plasma density, for low energy electrons in a relatively dense loop, energy lost by CC can be comparable to the RC losses and affect the dynamics of the beam/RC system. We derive here the equations which include CC self-consistently with RC losses.
 
 The energy loss equation of a nonthermal beam electron undergoing RC and collisional losses in a hydrogen atmosphere with degree of ionization $\alpha$ is: 
 
 \begin{equation}
 {\frac{dE}{dx}} = {-e \mathscr{E}_{RC}(x) - 2\pi e^4{\frac{n_e(x)}{E}}( \alpha  \Lambda_{ee} + (1-\alpha) \Lambda_{eH}) } 
 \label{eq:energy_loss}
 \end{equation}
 where $\Lambda_{ee}$ and $\Lambda_{eH}$ are the electron-electron Coulomb logarithm for a fully ionized plasma and the effective Coulomb logarithm for electron-hydrogen atom collisions, $\alpha={\frac{n_i}{n_i+n_n}}$ is the degree of ionization where $n_i$ and $n_n$ are the ionized and neutral hydrogen densities, respectively. The second term in Equation~\ref{eq:energy_loss} is the energy loss by Coulomb collisions using Equation 23a in \cite{1978ApJ...224..241E}. 
The beam current density is affected by both RC and CC losses: 
\begin{equation}
J_{b} =
\begin{cases}
eF_{e0} & \mbox {; } x \leq x_{C} \\
eF_{e0} \left(\frac{E_{min}(x)}{E_{c0}}\right)^{1-\delta}   & \mbox{; } x >x_{C} 
\end{cases}
\label{eq:flux}\end{equation}

where $x_{C}$ is the thermalization distance which is equivalent to $x_{RC}$ if CC losses are negligible. If RC losses are negligible, $x_{C}$ is defined as the distance where the energy lost by an electron with initial energy $E_{c0}$ becomes equal to $E_{th}$, i.e.,  
\begin{equation*} 
E_{th}^2-E_{c0}^2=-2\pi e^4 \int_0^{x_{C}} n(x')(\alpha\Lambda_{ee}+(1-\alpha)\Lambda_{eH})dx'.
\end{equation*} 

$E_{min}(x)$ is calculated numerically using Equation~\ref{eq:energy_loss} and the flux conservation equation $F(E,x)dE = F_0(E_0,x=0) dE_0$. When Coulomb collisional losses are negligible,
\begin{equation*}
E_{min}(x)=E_{th}(x)+e\int_0^x \mathscr{E}_{RC}(x') dx'.
\end{equation*} 
Similarly, if return current losses are negligible, 
\begin{multline*}
E_{min}(x)=  \Big( E_{th}^2(x) + \\4\pi e^4\int_0^x n_e(x')(\alpha\Lambda_{ee} +(1-\alpha)\Lambda_{eH})dx' \Big)^{1/2}.
\end{multline*}

The dimensionless current balance equation which includes RC and collisional losses, acceleration of runaway electrons, and thermalization of electrons is the following: 

\begin{align}  {\frac{eF_{e0}\eta_s(x)}{\mathscr{E}_D(x)}}\left( \frac{E_{min}(x)}{E_{c0}} \right)^{1-\delta}= z(x) - \nonumber \\ \frac{e \eta_s(x)}{\mathscr{E}_D(x)}\int_x^L n_e(x') \gamma_{run}(x')  dx' 
\label{eq:general}\end{align}
 
This equation generalizes Equation~\ref{eq:510} to include Coulomb losses. It is solved using the same method described in section~\ref{sec:RA.RC}, with the difference being that the initial guess for the beam current density is calculated numerically using Equation~\ref{eq:flux}.  We will refer to this model as RA.RC.CC, and the case without runaways as RC.CC where RC and CC losses are considered but $J_b=-J_d$.

We assume that the energy lost by the beam (Equation~\ref{eq:energy_loss}) is partially gained as heat by the ambient plasma and accelerates the remaining fraction. This is calculated self-consistently.
 So in this case, the volumetric energy flux lost by the nonthermal beam is \begin{equation}q(x)=-\frac{d}{dx}\int E F(E,x) dE \label{eq:heat1}\end{equation}
whereas the heating rate is \begin{equation}Q(x)=\mathscr{E}_{RC}^2/\eta_s \label{eq:heat} \end{equation}
Table~\ref{tab:models} summarizes the limiting cases and nomenclature used. The fourth column shows other papers with similar assumptions about the mechanisms taken into account.

 \begin{table}[ht!]
%\begin{center}
\begin{tabular}{c c c c}

Model & Beam & Current & Similar   \\
  & Losses &Balance &conditions\\
\hline \hline %\vspace{0.5cm}

 RA.RC.CC & RC\&CC  &$j_b=j_d+j_{run}$ & N/A \\

RA.RC & RC & $j_b=j_d+j_{run}$ & N/A\\

RC.CC & RC\&CC & $j_b=j_d$  & E80+K15, \\
& &  & ZG06  \\
& &  & A20 \\

RC.only & RC & $j_b=j_d$ & H12+K15,\\
 & & & A20 \\

CC.only & CC & N/A & E78+K15,\\
& & & A20 \\
\hline \hline
\end{tabular}
%\end{center}
\caption{Summary of models. The first column is the model name. The second column lists the physical mechanisms included in the model by which beam energy is lost. The third column lists the current balance equation implemented by the model. For the CC.only case, the return current is not modeled, so there is no current balance equation. E80 is ~\cite{1980ApJ...235.1055E}, K15 is \cite{2015ApJ...809...35K}, ZG06  is \cite{2006ApJ...651..553Z}, A20 \cite{2020ApJ...902...16A}, H12 is \cite{2012ApJ...745...52H}, and E78 is \cite{1978ApJ...224..241E}.}
\label{tab:models}
\end{table}

   \section{Application to solar flare loop modeling}
   \label{sec:application}
 \subsection{Spatial evolution of the current densities and RC electric field}
As a first example we choose a test demonstrating the case with energy loss less than the thermalization energy (\S\ref{sec:notherm}). We model the transport of an electron beam with high flux density propagating in a cool/less dense loop , in which CC losses are negligible. We inject a high flux density of beamed electrons to produce a large return current density (since $j_{RC}$ must balance $j_b$). We use a cool loop to increase the resistivity and therefore the electric field.  We will show that these conditions also imply that the effects of  runaway electrons are critical.
\begin{figure}[h!]
\centering
    \includegraphics[width=0.45\textwidth]{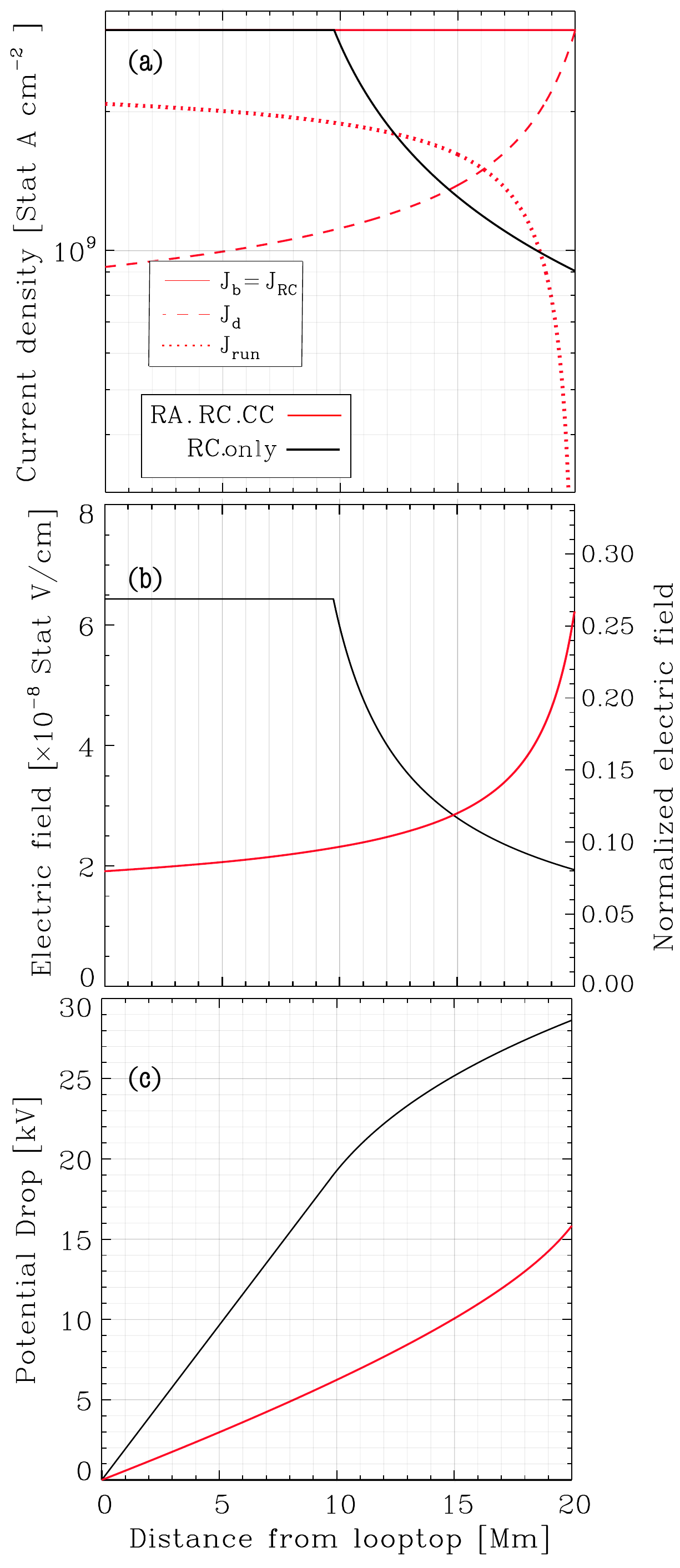}
   \caption{The current density (a), electric field (b), and potential drop (c) for the RA.RC.CC model (red lines) and the RC.only model (black lines) given an injected beam flux density of $6.25\times 10^{18}$ e$^-$cm $^{-2}$s$^{-1}$ (the (the corresponding energy flux density is $3\times 10^{11}$ erg cm$^{-2}$ s$^{-1}$ labeled,  3F11), E$_c=$20 keV and $\delta=4$ in a 20 Mm half-loop with temperature T=3.5 MK and density $n_e=4.5\times 10^{9}$ cm$^{-3}$. These beam/background plasma parameters will be referred to as case 1 throughout the paper.}
  \label{fig:example_with_without}
\end{figure} 

Figure~\ref{fig:example_with_without} shows examples of the predicted spatial evolution of return current densities, electric fields, and potential drops,  that include and exclude the effects of runaways as defined in \S\ref{sec:notherm}.  In each model, the injected electron beam is characterized by a power-law with $F_{e0}=6.25\times 10^{18}$ e$^-$cm $^{-2}$~s$^{-1}$, E$_c=$20 keV and $\delta=4$ propagating in a 20 Mm half-loop with apex temperature, T = 3.5 MK, and apex density, $n_e=4.5\times 10^{9}$ cm$^{-3}$. These beam and background parameters will be referred to as case 1 in the remainder of the paper. In each panel the red and black lines correspond to the RA.RC.CC and RC.only models, respectively. 

 The sudden decreases in current density (panel a) and electric field (panel b) in the RC.only model are due to the thermalization of electrons out of the beam occurring at about 10 Mm from the looptop. Only the drifting component of the return current density is Ohmic, so the inclusion of runaways results in an electric field that is \emph{less} than that predicted by RC.only in most of the upper part of the loop (about three-quarters of the loop length). In fact, in this example the electric field produces a total potential drop that is smaller than the thermalization energy, so in the RA model no electrons are lost from the beam throughout their transport.  The runaway current increases from the footpoints to the looptop as the electric field accelerates more background electrons, and the drifting current is reduced as a consequence of the current balance assumption. Notice that the runaway current density dominates the drift density over much of the loop. At the loop top the runaway density is 69$\%$ of the total RC density.

Finally notice the RC electric field is lower than $0.26\mathscr{E}_D$ everywhere along the loop and $\sim0.1\mathscr{E}_D$ in the upper corona.

This example shows that for a beam with injected flux density of 6.25$\times10^{18}$ cm$^{-2}$ s$^{-1}$ propagating in a cool, relatively low-density coronal loop, the runaway electrons significantly change the dynamics of the direct beam/RC system by reducing the local electric field.

\begin{figure}[h!]
\centering
   \includegraphics[width=0.47\textwidth]{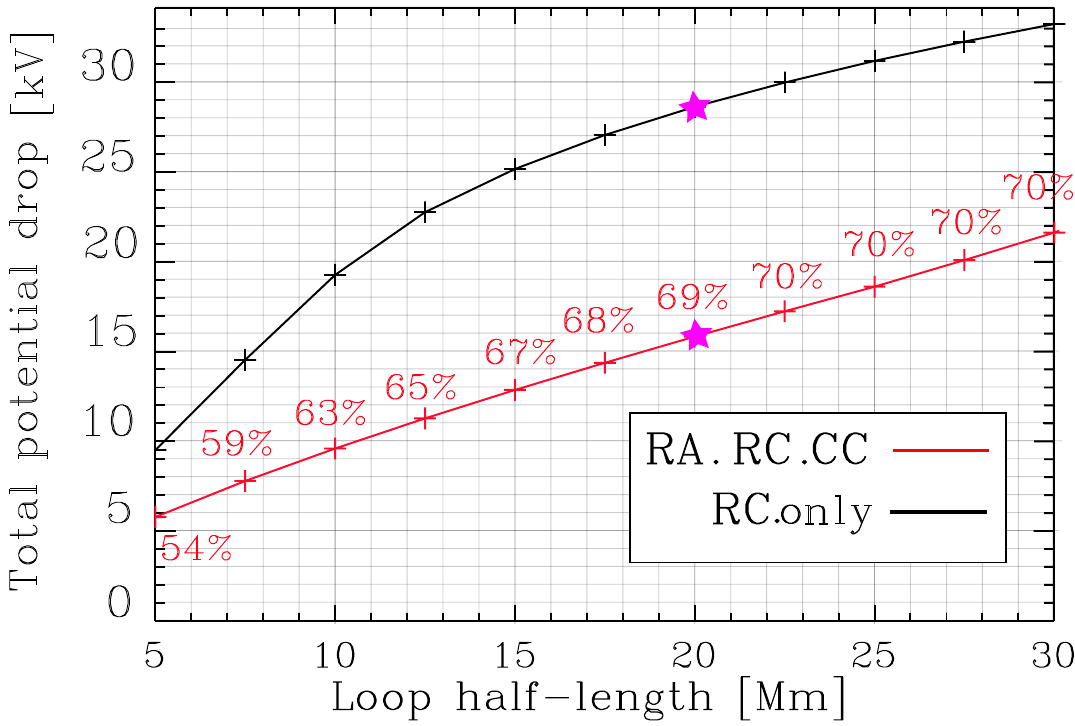}
   \caption{Potential drop from the looptop to the footpoints as a function of loop half-length with temperature and apex density given by $T = 3.5$~MK, and $n_e=4.5\times 10^{9}$~cm$^{-3}$, respectively. The injected flux density is $F_{e0}=6.25\times10^{18}$ cm$^{-2}$ s$^{-1}$, $\delta=4$, E$_c=20$ keV. The percentages correspond to the fraction of the RC flux carried by runaway electrons at the looptop. The stars show the solutions from Figure~\ref{fig:example_with_without} (case 1).}
  \label{fig:potential_length}
\end{figure} 

For the next test case, we have varied the half-length of our model loop. Of course, the longer the distance the beam has to propagate, the higher the resulting potential drop. For this case, we use the same injected electron distribution into loops with the same temperature and density as was used in Figure~\ref{fig:example_with_without}. The results of this experiment are shown in Figure~\ref{fig:potential_length}, which plots the potential drops for the RA.RC.CC and RC.only models as a function of loop half-length. 

 In the RC.only case, the change in potential drop versus loop length slows for loops longer than about $10$~Mm. At that height the energy lost due to the potential drop is equal to the thermalization energy, so the beam current and hence return current densities lessen. Note that even though we have considered a maximum loop half-length of $30$~Mm, the loop half-length as determined from RHESSI observations is up to 37~Mm (AH17). 
 
 The fraction of the RC flux carried by runaway electrons saturates for loop lengths above 20~Mm. This is because no beam electrons are thermalized for loop lengths less than 20~Mm and as beam electrons have to propagate along longer loops, the potential drop increases resulting in thermalization of the lower energy electrons. The longer the loop, the higher (closer to the looptop) the thermalization distance in both models with and without runaways. However, in the runaway case, the potential drop gained by traveling along a longer loop is compensated for by the loss of electrons from the beam.

\begin{figure}[h!]
\centering
  \includegraphics[width=0.46\textwidth]{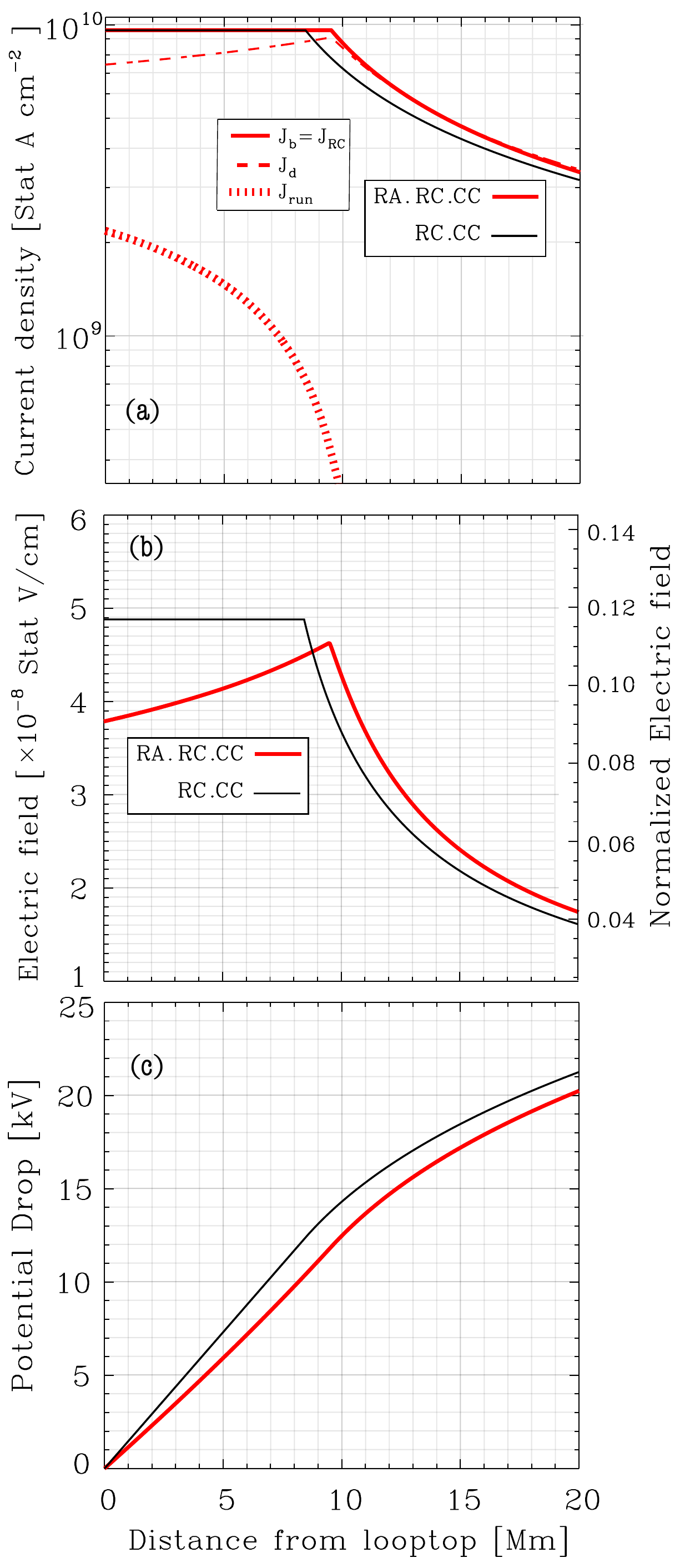}
   \caption{Same as Figure~\ref{fig:example_with_without} using T=9 MK and $n_e=2\times10^{10}$ cm$^{-3}$, and an injected electron flux density $F_{e0}=2\times10^{19}$ cm$^{-2}$ s$^{-1}$, energy flux density $\mathscr{F}_0=1\times10^{12}$ erg cm$^{-2}$ s$^{-1}$. The black curves are calculated using the RC.CC model. These beam/background plasma parameters will be referred to as case 2 throughout the paper.}
  \label{fig:thermal}
\end{figure} 
As a final experiment in this section, we compare the RA.RC.CC and RC.CC models. We have constructed this experiment to demonstrate a case in which beamed electron thermalization begins to occur in the coronal portion of the loop (i.e., applicable to the theories in \S\ref{sec:RA.RC} and \S\ref{sec:RA.RC.CC}) rather than the footpoint (as was the case in Figure~\ref{fig:example_with_without}). We inject an electron beam characterized by flux density, power-law index, and cutoff energy given by, $F_{e0}=2\times10^{19}$ cm$^{-2}$ s$^{-1}$, $\delta = 4$ and $E_{c0}=20$ keV, respectively. The corresponding injected energy flux density is $1 \times 10^{12}$ erg cm$^{-2}$ s$^{-1}$. The beam is injected into a $20$~Mm half-length loop with apex temperature and density of $T=9$~MK $n_e = 2\times10^{10}$ cm$^{-3}$, respectively. These beam and background parameters will be referred to as case 2 in the remainder of the paper. The results of this experiment are shown in Figure~\ref{fig:thermal}. In the RA.RC.CC case, the electric field reaches a maximum at the position where electron thermalization begins to occur, $x = 9.4$~Mm. As electrons are thermalized, the beam and return current densities and therefore electric field begin to drop.

\subsection{What fraction of suprathermal electrons return to the acceleration region?}  
The local contribution to the runaway electrons, $\gamma_{run}$, (most easily inferred from the analytical solution of Equation~\ref{eq:kb64}), depends on the normalized electric field and the collision frequency. In some cases the local contribution of runaways is a small fraction of the local drifting Maxwellian, but the cumulative buildup of runaways along the loop means that  as they reach the looptop, their total contribution may be substantial. This is the case for the experiment plotted in Figure~\ref{fig:example_with_without}. In other circumstances, the local runaway contribution may become a significant fraction of local Maxwellian distribution (i.e., where the square of the ratio of the thermal speed to the critical speed is not much less than one, resulting in $\mathscr{E}_{RC}/\mathscr{E}_D$ close to one). In these cases, the assumption to treat the drifting and runaway components as separate distributions breaks down. The return current density, $J_{RC}$, should be calculated as a moment of a combined distribution. Since the runaway growth rate also depends on the collision frequency (Equation~\ref{eq:freq}), for lower temperature and/or higher density, more runaway electrons are accelerated locally for the same normalized electric field strength. An example demonstrating how the electron distribution differs from a Maxwellian as a function of normalized electric field strength is shown Figure~3 in \citet[][]{2017CoPhC.212..269S}.
 
It is important to recognize the difference between the local and cumulative fraction of runaway electrons. The local accelerated value depends on the local density, temperature and magnitude of the electric field, whereas the fraction of runaway electrons at the looptop is the sum of all accelerated runaways. By the same token, the minimum energy of electrons in the runaway population at position $x$ that were accelerated out of thermal distribution at position $L$ is 
\begin{equation}
E_{*}(x)=\frac{m}{2}v_{cr}^2(x)+ \int_L^x e\mathscr{E}_{RC}(x')dx',
\label{eq:E_crit}
\end{equation} 
where $v_{cr}$ is the critical velocity in Equation~\ref{eq:crit_vel}. 
$E_{*}(x)$ depends on how the electric field and the Dreicer field vary along the loop.
The critical energy is the lowest energy of runaway electrons.
For the examples shown in Figures~\ref{fig:example_with_without} and ~\ref{fig:thermal}, the minimum runaway energies at the looptop are given by $E_{*}(0) = 16$~keV and $E_{*}(0) = 21$~keV, respectively. These energies are of the same order of magnitude as the accelerated nonthermal beam electrons. For comparison, the thermal energy in the first and second cases are 0.3 keV and 0.8 keV.

For the experiment in Figure~\ref{fig:thermal}, notice that even though the total potential drops from the looptop to the footpoints are similar in the RA.RC.CC and RC.CC models (20 kV and 21 kV, respectively), the nature of the electrons at the looptop (and assumed to be returning to the acceleration region) is significantly different. In the RA.RC.CC model, 22.7$\%$ of the electrons returning to the acceleration region are suprathermal with a minimum energy of 21 keV. Note that the assumption that electrons return to the acceleration region is reasonable if the time to reach the steady-state is lower than the acceleration time. The time to reach the steady-state in atmospheres 1 and 2 are 21~ms and 20~ms, respectively.

The energy gained by the runaway electrons at the looptop is equal to the total potential drop and is just over 20 keV (second term in Equation~\ref{eq:E_crit}). The minimum energy in the runaway current distribution depend on the values of $E_{cr}$ and the actual shape of the drift current electron distribution.  Given that the electrons start with an energy of at least 0.8 keV, 21.8 keV is the energy of the ``bulk" of the runaway electron distribution at the top of the loop.

\subsection{Model comparison of heating rates}
\label{sec:heating}
\begin{figure*}[bth!]
\centering
    \includegraphics[width=0.95\textwidth]{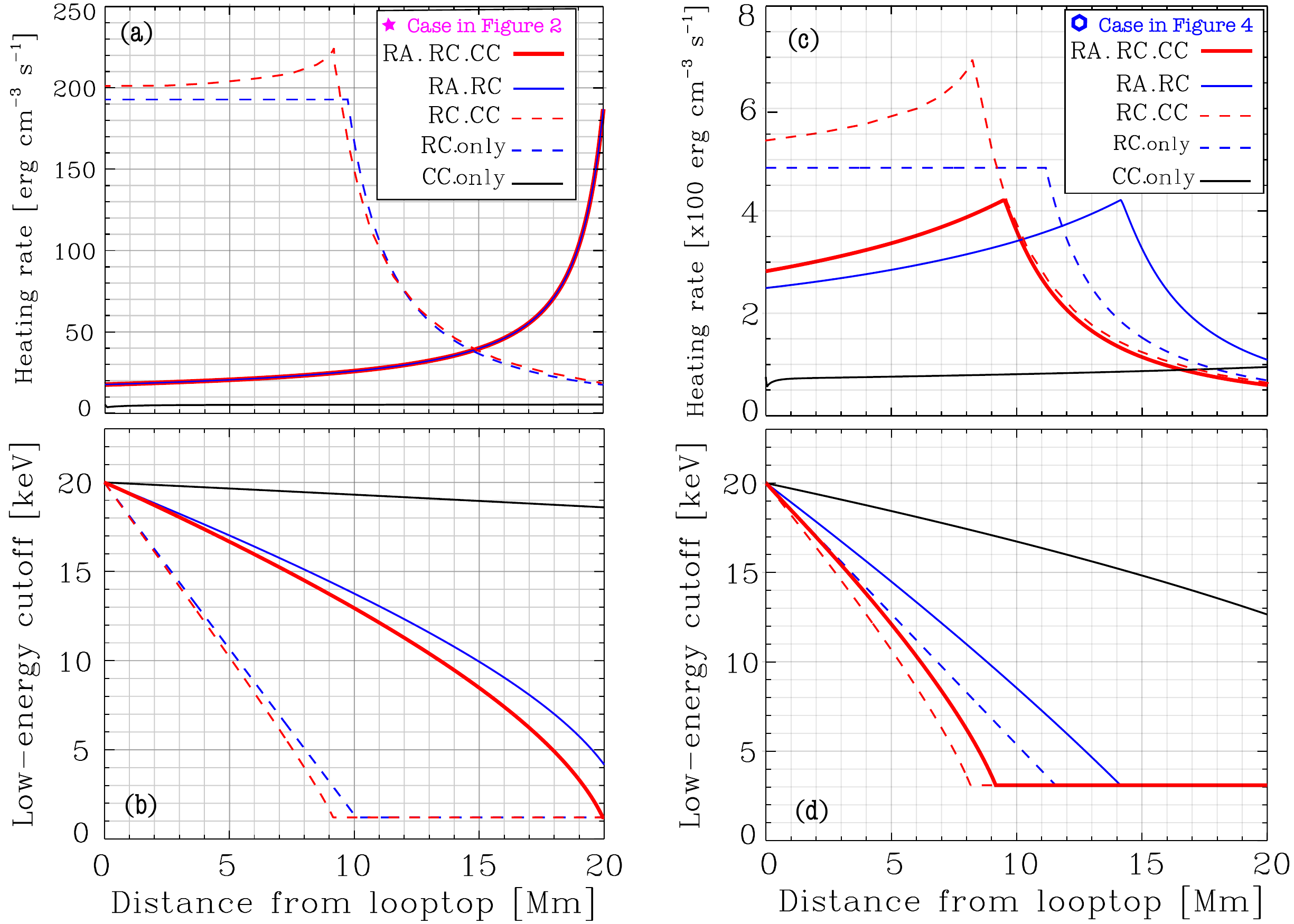}
   \caption{Comparisons of the volumetric heating rates (a,~c) and low-energy cutoffs (b,~d) as a function of distance computed by the five models listed in Table~\ref{tab:models}. The left side panels (a,~b) are for an injected electron beam and loop that are identical to those used in the example shown in Figure~\ref{fig:example_with_without}. The right side panels (c,~d) are for an injected electron beam and loop identical to those used in Figure~\ref{fig:thermal}.}
  \label{fig:energy_loss_comp}
\end{figure*}
Here we include some more detailed comparisons of results from our analytical (CC.only and RC.only) and numerical (RC.CC, RA.RC and RA.RC.CC) models. In Figure~\ref{fig:energy_loss_comp}~a,b, we show the volumetric heating rates and low-energy cutoffs as functions of position for our five models listed in Table~\ref{tab:models}. The loop half-length, apex temperature and density, and injected beam distribution are identical to those used in the experiment presented in Figure~\ref{fig:example_with_without} (case 1). Figure~\ref{fig:energy_loss_comp}~c,d show similar quantities for loop conditions and injected beam distribution identical to the case presented in Figure~\ref{fig:thermal} (case 2).  The case in the left panels is chosen to illustrate conditions where energy losses from the return current are significant but Coulomb collisions losses are negligible. In the right panels, return current losses, Coulomb collisions and runaways, all significantly affect the solution. 

Although runaway electrons reduce Joule heating overall, panel a in Figure~\ref{fig:energy_loss_comp} shows that, in the models that exclude runaways, the heating in the upper corona is overestimated by an order of magnitude while in the lower corona the heating rate is underestimated by as much as an order of magnitude when compared to models that include runaways. In panel b, for the RA.RC.CC and RA.RC models, the total potential drop is reduced enough by the presence of runaways for all electrons to reach the thick target without being thermalized. In contrast, in the RC.CC and RC.only models, the beam flux loses $\sim70\%$ of its electrons to thermalization before it reaches the footpoints (panel a of Figure~\ref{fig:example_with_without}). This results in a factor of 11 decrease in the heating rate in the lower corona (panel a of Figure~\ref{fig:energy_loss_comp}). As already mentioned, Coulomb collisional losses are negligible in this example, as evidenced by little change in the low-energy cutoff and nearly zero heating for the CC.only case. But they do have create a small increase of the heating rate of RC.CC compared to RC.only.

For the case presented in panels c and d, the heating rate is lower by a only a factor 1.8 in the upper corona in the runaway case (RA.RC.CC) compared to the model without runaways (RC.CC). As expected, the thermalization distance is lowest when RCs and CCs are considered without runaways. The thermalization distance in the RA.RC.CC model is lower down the loop because RC losses dominate over CCs even when runaways are accounted for. For the models excluding CC losses (RA.RC and RC.only), the thermalization distance is lower in the case with highest energy losses, i.e., without runaways. Finally, notice that since the density is higher in loop used in the right panels compared that used in the left panels, the heating by CC losses is more significant. This is demonstrated in panel~d, for the CC.only case an electron with initial energy of 20~keV loses 7~keV before it reaches the footpoint.     
These examples show that runaway electrons decrease the heating rate compared to collisional return current models (RC.CC and RC.only), but the heating is significantly higher than models neglecting RC losses altogether. We chose relatively high but reasonable injected flux densities (as inferred by spectral fitting of X-rays observed during solar flares) to demonstrate the effect of runaway electrons. In the next section, we show how these cases compare to other injected beam flux densities and coronal atmospheres.  

\begin{figure}[bth!]
\centering
    \includegraphics[width=0.38\textwidth]{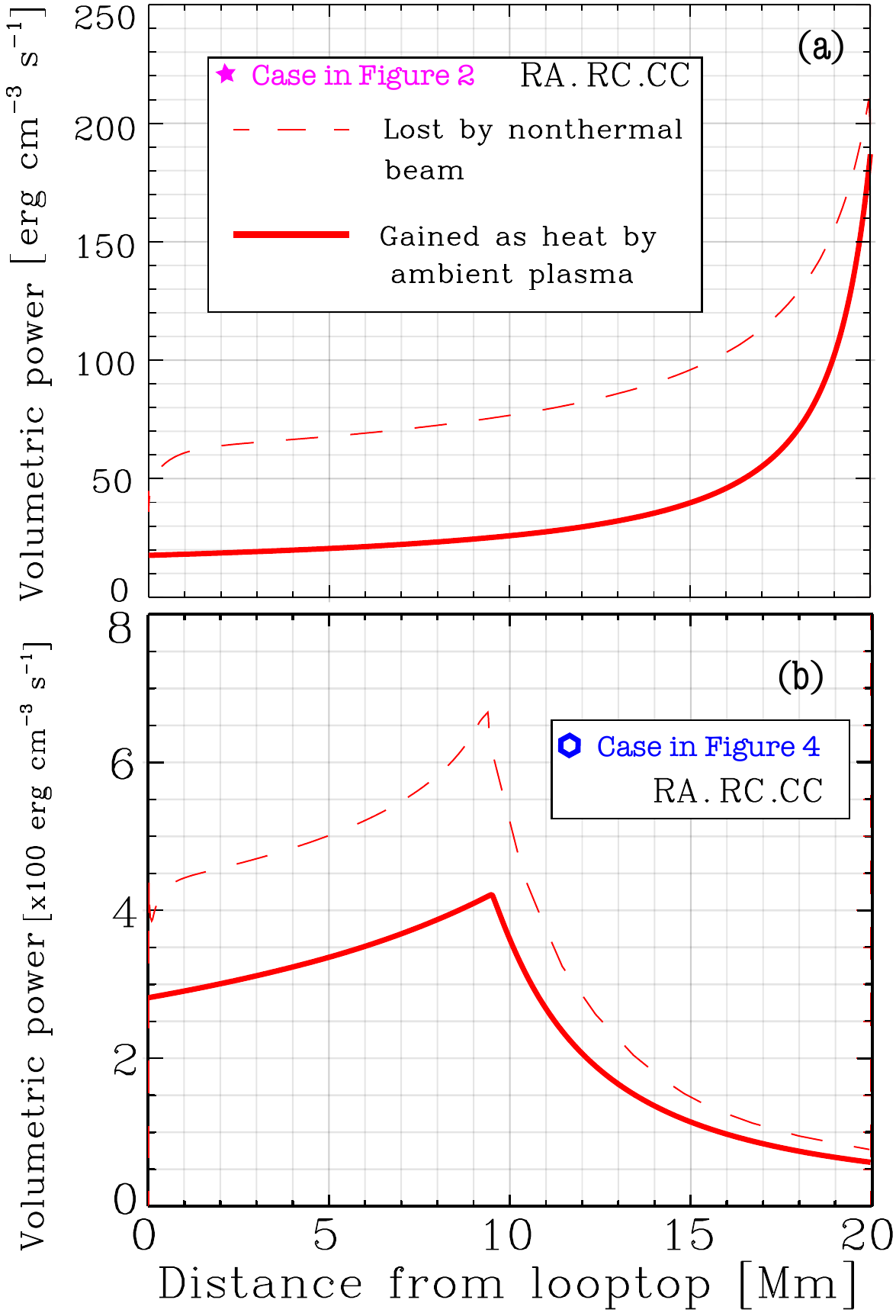}
   \caption{Comparisons of the volumetric power lost by the nonthermal beam (dashed) and gained by the ambient plasma as heat (solid). The full model RA.RC.CC was used for these calculations. The solid red curves are the same as the solid red curves in Figure~\ref{fig:energy_loss_comp}.}
  \label{fig:energy_loss_gain}
\end{figure}

Finally, Figure~\ref{fig:energy_loss_gain} shows the total volumetric power lost by the nonthermal beam as calculated by Equation~\ref{eq:heat1} and the volumetric power gained as heat by the ambient plasma as calculated by Equation~\ref{eq:heat}. The remaining energy lost by the beam served to accelerate ambient electrons into runaways. Notice that the total energy lost by the nonthermal beam in the runaway model is lower than that lost by the beam in the model without runaways (RC.CC, dashed red curves in panels a and c of Figure~\ref{fig:energy_loss_comp}).

\section{Beam propagation in various uniform coronal atmospheres}
\label{sec:var_atm}
Let us now consider the injection of electron beams into five loops with different atmospheric temperatures and densities, but each with a loop half-length of 20 Mm. A range of injected electron flux densities between $10^{17}-10^{20}$ e$^-$ cm$^{-2}$ s$^{-1}$ is considered, with a spectral index $\delta=4$, and two values of the low-energy cutoff at 20 and 50 keV. Table~\ref{tab:atm} lists the temperature, density and corresponding Dreicer field magnitude of the five atmospheres. For each loop, we assume constant temperature and density to isolate the effect of runaway electrons from those due to the background plasma parameters variation. We note that runaway electrons are more significant in the corona because the Dreicer field in the chromosphere is much higher than in the corona as the density is higher and the temperature lower in the chromosphere. Therefore, the fraction of runaways built up in the chromosphere can be neglected.

\begin{table}[ht!]
\begin{tabular}{c c c c}
Atmosphere& T  & Density & $\mathscr{E}_D$  \\
number& [MK]&[cm$^{-3}$]  & [StatV cm$^{-1}$]\\
\hline \hline 

 1 &3.5 &$4.5\times10^9$ & $2.4\times10^{-7}$ \\

2& 9 &$2\times10^{10}$ & $4.2\times10^{-7}$\\

3 & 20 &$5\times10^{10}$ & $4.8\times10^{-7}$\\

4 &35 &$1\times10^{11}$ & $5.5\times10^{-7}$ \\

5 & 20 & $1\times 10^{10}$ & $1.0\times10^{-8}$\\
\hline \hline
\end{tabular}
%\end{center}
\caption{Atmospheres used in Figure~\ref{fig:example}. The second and third columns are the electron temperature and density, respectively. The fourth column is the corresponding value of the Dreicer field. The first two atmospheres correspond to those used in the examples in Figures~\ref{fig:example_with_without}~and~\ref{fig:thermal}, respectively.}
\label{tab:atm}
\end{table}

\begin{figure}[hb!]
\centering
   \includegraphics[width=0.39\textwidth]{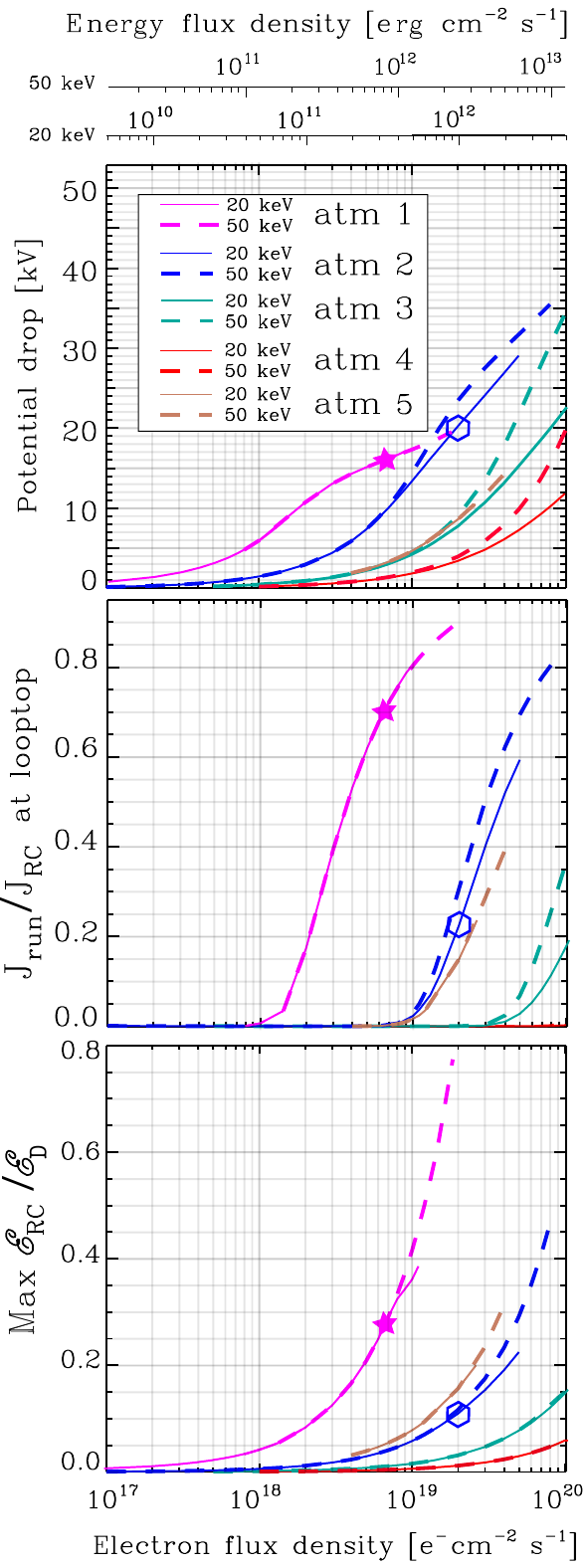}
   \caption{Potential drop (top), runaway current fraction (middle), and maximum RC electric field normalized by the Dreicer field (bottom). For the injection of electrons with beam flux densities between 10$^{17}$ and 10$^{20}$ cm$^{-2}$ s$^{-1}$, and spectral index $\delta=4$ into the atmospheres listed in Table~\ref{tab:atm}. The solid and dashed curves represent low-energy cutoffs of $E_{c}=$20 and 50 keV, respectively. The star and polygon are the solutions for the cases presented in Figures~\ref{fig:example_with_without}~and~\ref{fig:thermal}, respectively.}
  \label{fig:example}
\end{figure}

Our most complete model, RA.RC.CC, is used in this section. We will discuss how the potential drop and the runaway fraction of the RC flux at the looptop depends on the nonthermal beam parameters and the temperature and density of the background plasma. Figure~\ref{fig:example} shows the results of the electron beam injection into the five atmospheric models listed in Table~\ref{tab:atm}. The upper panel shows that nonthermal electrons with a higher injected flux density induces a higher RC electric field and therefore a higher potential drop between the looptop and footpoints. This trend does not depend on the atmosphere in which the beam is streaming. However, the atmosphere with the lowest temperature (highest resistivity) results in a higher induced electric field, for the same injected beam parameters.  

The highest injected flux density in each atmosphere satisfies the condition that the beam density, $n_{b0}$, is no more than 30\% of the background coronal density. In Atmospheres 3 and 4, the highest injected densities are lower. This is a convenient way to keep the RC electric field low enough compared to the Dreicer field. We will discuss in \S\ref{sec:instability} whether this condition is sufficient to keep the beam stable to current-driven instabilities.

The solutions are sensitive to the low-energy cutoff value $E_{c0}$ because it takes a lower electric field strength and potential drop to thermalize lower energy electrons. In addition, lower energy electrons lose more energy by CC losses which also results in shortening the distance of thermalization, $x_{C}$. This explains differences between the cases with low-energy cutoff of 20~keV and 50~keV (solid and dashed lines, respectively). The potential drop is therefore independent of the low-energy cutoff if it is lower than $E_{c0}-E_{th}$. However, if any non-thermal beam electrons are thermalized, the potential drop is lower for the beam with the lower low-energy cutoff. This happens because the thermalization lowers the electron flux and electric field, and because the thermalization distance is further lowered by CC losses which are more significant for 20~keV electrons compared to 50~keV electrons.

Injecting two beams with the same total electron flux density but different low-energy cutoffs into a plasma results in comparable potential drops if none of the electrons are thermalized. This is useful to understand the behavior of the return current, but these two beams do not carry the same energy and they are observationally distinct, as the beam with a low-energy cutoff of 50 keV results in a flattening of X-ray spectra at least at 50 keV. The X-ray spectrum is further flattened at lower energies by the potential drop \citep[e.g.,][]{2012ApJ...745...52H}. 
 
The lower panel shows that the RC electric field is sub-Dreicer for all the solutions. The location of the maximum electric field strength can be anywhere along the loop and corresponds to where beamed electrons begin to be thermalized, or at the footpoints if thermalization begins there. 

The minimum injected flux density to accelerate significant runaway electrons is higher for beams propagating in hotter atmospheres. A higher electric field is needed to accelerate electrons out of a hotter and/or less dense atmosphere because the dynamic friction force is lower. The dynamic friction force peaks at $v\sim v_e$ with a value of $e\mathscr{E}_D\propto \frac{n_e}{T}$ \citep[e.g.,][]{1996RaSc...31.1541G,2012ASSL..391.....S}. Atmospheres 3 and 5 have the same temperature, T=20 MK, and different densities, $n_e=5\times10^{10}$ cm$^{-3}$ and $n_e=1\times10^{10}$ cm$^{-3}$, respectively. Atmosphere 3, having the highest density, accelerates a smaller fraction of runaway electrons.

\subsection{Self-reducing effect of runaways}

\label{sec:self-reduce}
Interestingly, the influence of runaway electrons on the RC electric field is self-reducing. A higher injected flux produces a higher electric field, resulting in a higher rate of runaways. But a higher contribution of runaways \emph{decreases} the RC electric field, thereby decreasing the runaway rate. Thus, the solution is found where the self-regulating effect is balanced. This can be seen in the upper panel of Figure~\ref{fig:example}. Increasing the injected electron flux density results in the eventual decrease in electric field between the looptop and footpoints if no beam electrons are thermalized. However, if thermalization occurs, the potential drop between the looptop and footpoints is reduced but the local electric field induced by the lower injected flux density beam is higher below the thermalization distance. For the Atmosphere 1 case, this result is seen for beam fluxes above $\sim 2 \times 10^{18}$ electrons cm$^{-2}$ s$^{-1}$, and in the Atmosphere 2 case above $\sim1.3\times 10^{19}$ electrons cm$^{-2}$ s$^{-1}$ (dashed blue curve). 

Another self-reducing effect is caused by the thermalization of beamed electrons. A larger beamed electron flux produces a stronger electric field, which thermalizes more electrons out of the beam, thereby reducing the electric field. This can be inferred by differences in the potential drops between the $E_{c0} = 20$~keV and $E_{c0} = 50$~keV cases that are plotted in the top panel of Figure~\ref{fig:example}. For the 20~keV case electrons are thermalized for lower injected flux densities, whereas in the 50 keV case many beam electrons reach the footpoints without thermalization, even for the highest injected fluxes considered. Increasing the injected flux density, moves the thermalization distance closer to the injection site, so the beam flux density decreases producing a weaker electric field.

\section{Effects of pitch-angle and instabilities}
\label{sec:discussion}
\subsection{Beam electrons versus runaway suprathermal electrons returning to the looptop}
\label{sec:pitch-angle}
\begin{figure*}[bth!]
\centering
    \includegraphics[width=0.99\textwidth]{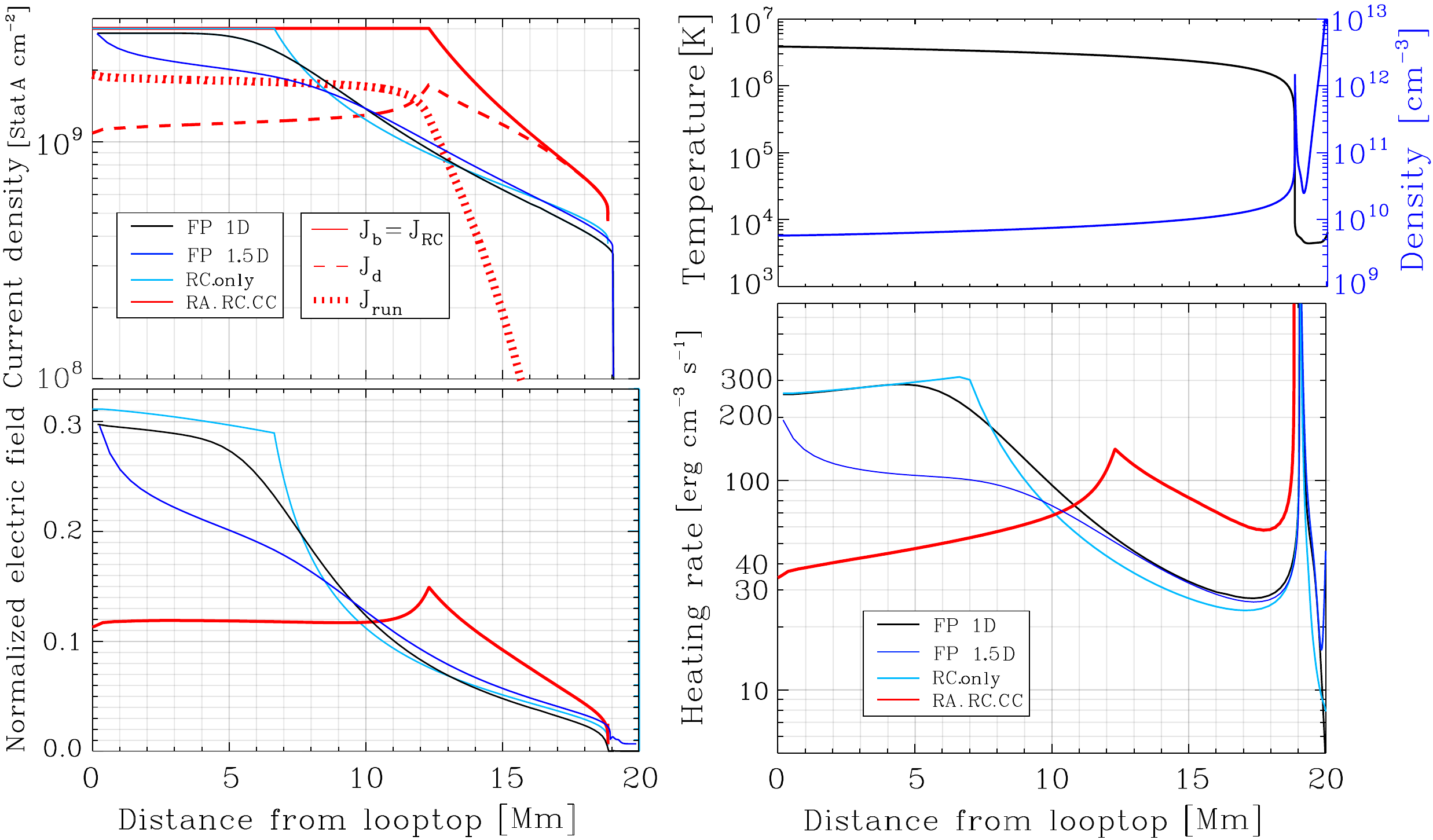}
   
   \caption{Comparison of the RA.RC.CC model with results from H12+K15 and A20. FP~1.5D and FP~1D label results from the FP model configured to include or exclude changing pitch-angles, respectively. Top left: Beam current density and equivalently RC density from the FP~1D (black), FP~1.5D (blue), RC.only (cyan), and RA.RC.CC (solid red) models. For the RA.RC.CC model, the current density has been subdivided into drifting (dotted red) and runaway (dashed red) components. Bottom left: Normalized RC electric field. Top right: Loop electron temperature (black) and density (blue) stratification. Bottom right: Heating rates from the four models. For the bottom panels, line colors refer to the same models as the top left panel.}
  \label{fig:comparison}
\end{figure*} 

In this section, we discuss the relative effect of our 1D assumption (i.e., assuming all electrons move parallel to the magnetic field) compared to a 1.5D assumption in which electrons may have non-zero pitch angles.  We aim to compare the fraction of runaways at the looptop (using RA.RC.CC) to the reversed nonthermal beam electrons (using FP 1.5D RC\&CC). 

Because of the RC electric field, initially beamed electrons systematically lose the component of their energy parallel to the magnetic field. They are decelerated and turned backward if the potential drop is greater than their parallel energy.  Additionally, Coulomb collisions scatter these electrons in all directions, producing a broader pitch-angle distribution. Scattering, the reversal due to deceleration, and the re-acceleration by the RC electric field are significant. These reversed electrons are effectively part of the return current, reducing the needed balancing drifting component, the RC electric field, and RC losses.

To compare the electron beam transport in 1.5D with our 1D model, we use the Fokker-Planck kinetic transport model, FP, developed by A20. FP models the distribution of nonthermal electrons injected at the looptop as they transport to the footpoints, under the influence of Coulomb collisions and pitch-angle scattering and the return current electric field force, but without considering the runaway component. Figure~\ref{fig:comparison} shows an example of this comparison. The loop temperature and density stratification are shown in the upper right panel \citep[][]{2005ApJ...630..573A,2015ApJ...809..104A}. We label FP configured to include or exclude forces changing the pitch-angle as FP~1.5D and FP~1D, respectively. First notice the similarities in current density, electric field and heating rate of the 1D models excluding runaways (RC.only and FP~1D). Since, FP~1D models the CC force and RC.only does not, thermalization is predicted to occur higher in the corona in FP~1D than in RC.only. The heating rate rises slowly over the first $\sim5$ Mm because the decreasing temperature results in increased resistivity. The decrease in heating rate beginning between 5 and 7 Mm is due to thermalization of the beam electrons. The decrease is smoother in the FP~1D case because energy diffusion is considered and treated self-consistently. 

Since FP~1.5D models the evolution of the pitch-angle, $\mu$, it is necessary to specify an injected pitch-angle distribution. To best compare with our 1D models, in FP~1.5D we use a distribution with all injected electrons having zero pitch-angle (i.e., $\mu_0=0$). The heating rate predicted by FP~1.5D is dominated by RC losses but is reduced by a factor 1.4 at the looptop and up to 2.8 at 4.7 Mm compared to the 1D case because the pitch-angle scattering and subsequent potential drop reduce the mean downward velocity of the beam electrons. The initial sudden drop in current density and heating rate between 0 and 2 Mm is due to collisional scattering and the reduction of the heating rate for distances just greater than $2$~Mm is due to the deceleration by the RC electric field. 

\begin{figure}[h]
\centering
\includegraphics[width=0.47\textwidth]{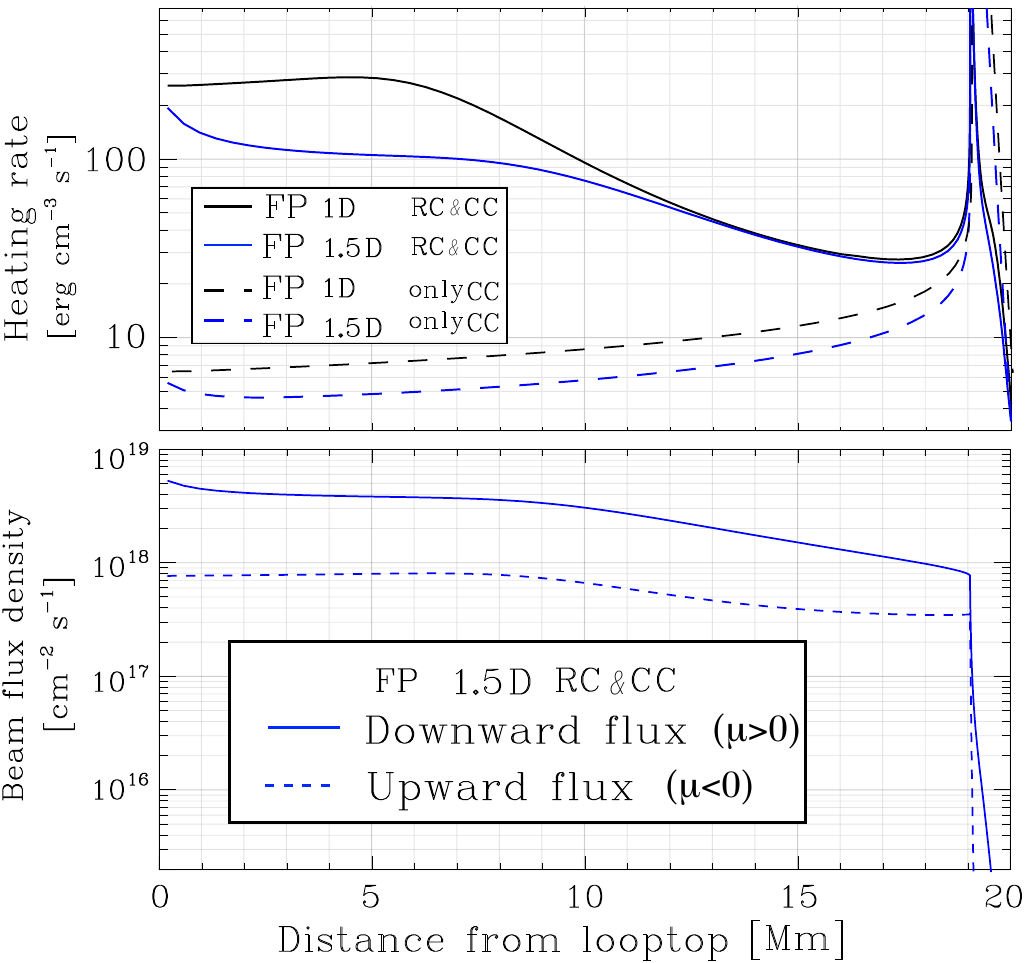}
   \caption{Upper panel: Comparison of heating rates using 1D FP, 1.5D FP with return current and collisional forces (solid blue), and collisional losses without return currents (dashed). Lower panel: Downward and upward propagating beam electrons ($\mu>0$ and $\mu<0$) in solid and dashed blue, respectively.}
\label{fig:beamed_comp}
\end{figure}

The upper panel Figure~\ref{fig:beamed_comp} shows the FP heating rates of the 1.5D and 1D models, in blue and black, respectively. The solid and dashed curves include both RC and CC forces, and only CC frictional forces, respectively. Notice the decrease in heating rate within 1~Mm of the looptop in both 1.5D solutions, with and without return currents, as evidence for the scattering being responsible for the initial drop in current density and heating rate from Figure~\ref{fig:comparison}. The lower panel of Figure~\ref{fig:beamed_comp}, shows the downward and upward propagating electrons. The latter are 14$\%$ of total RC flux returning to the looptop.

The current density in FP~1.5D is reduced compared to the FP~1D case (upper left panel of Figure~\ref{fig:comparison}). The initial scattering of electrons reduces the downward nonthermal beam flux, which results in a lower RC electric field compared to the 1D case. For the same reason the thermalization of beam electrons starts farther from the looptop. The difference in thermalization distance is also affected by the fact that electrons with non-zero pitch-angles lose less energy by the RC deceleration. 

Finally, we find that since the RC electric field predicted by the RA.RC.CC model is smaller throughout the top half of the loop compared to all other models, in this case runaways more effectively reduce the electric field than back-scattering or thermalization. The heating rate predicted by RA.RC.CC is lower than FP~1D and RC.only by almost an order of magnitude (a factor 0.13) in the upper corona. Additionally, the fraction of the RC in the runaway regime at the looptop is 64$\%$ (as inferred from the upper left panel of Figure~\ref{fig:comparison}) compared to 14$\%$ of the injected nonthermal electrons becoming reversed as predicted by FP~1.5D (lower panel Figure~\ref{fig:beamed_comp}). The fraction of reversed electrons depends on the injected pitch-angle distribution as well, and has been studied by \citet{1995A&A...304..284Z,2010A&A...512A...8Z} , and explicitly recognized by \citet{2009A&A...504.1057S} . The total potential drop in the corona is 33 kV in RA.RC.CC versus 106 kV in FP~1.5D. The current density is constant from the looptop to the thermalization distance in all 1D models (RA.RC.CC, RC.only, and FP~1D). 

The reversal of nonthermal beam electrons by the RC electric field and collisions was also studied by \citet[][]{1993SoPh..145..137K}, who recognized that both reversed electrons and runaways should be studied self-consistently but focused on the reversed component. He found that these electrons are higher by an order of magnitude compared to the runaway electrons accelerated out of the background plasma distribution. However, his estimate of the runaways density is simply the fractional part of the total density in the runaway tail \citep[][]{1978A&A....68..145N}. As shown by our model however, the runaways are accelerated and accumulate along the full length of the loop, and the RC electric field evolves as a function of the total runaway density.

Although runaway electrons can dominate the return current, their number will be affected by the reversed electrons as they both contribute to reducing the effect of return currents and the heating rates. While both contribute suprathermal electrons returning to the looptop, the number returning and the reduction of the RC electric field is larger in our runaway model. These are non-linear effects however, so they should be studied self-consistently.

In addition, two problems arise for $\mathscr{E}_{RC}\lesssim\mathscr{E}_D$  as opposed to $\mathscr{E}_{RC}\ll \mathscr{E}_D$ when considering 1D versus 1.5D treatments: (1) the velocity diffusion becomes non-negligible \citep{2008PhPl...15g2502S}, because the critical velocity of electrons approaches the thermal velocity, so that a 1.5D model is more appropriate. In fact the critical velocity is no longer a well-defined value for higher electric fields. (2) The runaway electrons can move into the slide-away regime \citep[][]{2016JPhCS.775a2013S}. This is a consequence of locally depleting the "bulk" distribution which effectively reduces the density, and therefore increases the effective Dreicer field. The result is that the entire background distribution is freely  accelerated even though $\mathscr{E}_{RC}<\mathscr{E}_D$. We will extend the model to 1.5 D (2D in velocity space) in an upcoming paper.

\subsection{Instabilities}
\label{sec:instability}
 Langmuir wave-beam instabilities can flatten the electron distribution at lower energies \citep[][]{2009ApJ...707L..45H} on time scales orders of magnitude lower than the collisional time scales, which are relevant for the beam/RC dynamics of interest in this paper \citep[][]{2011SSRv..159..107H,2011ApJ...733...33Z,2012A&A...544A.148K}. Other beam instabilities can also affect return current losses such as the Weibel instability, which can isotropize the beam if the magnitude of the guiding magnetic field is weak \citep[e.g.,][]{2009A&A...506.1437K}. A more comprehensive simulation is needed to understand the effect of beam instabilities and runaway electrons on the beam/RC system.
 
Rather, this section addresses the question of whether the drift speed is sufficiently low so that the beam/RC system is stable to parallel current-driven instabilities. The goal is to check whether these instabilities may arise under the conditions of our runaway model and discuss the considerations involved in analyzing the nonthermal beam/RC system.

\begin{figure}[h!]
\centering
   \includegraphics[width=0.45\textwidth]{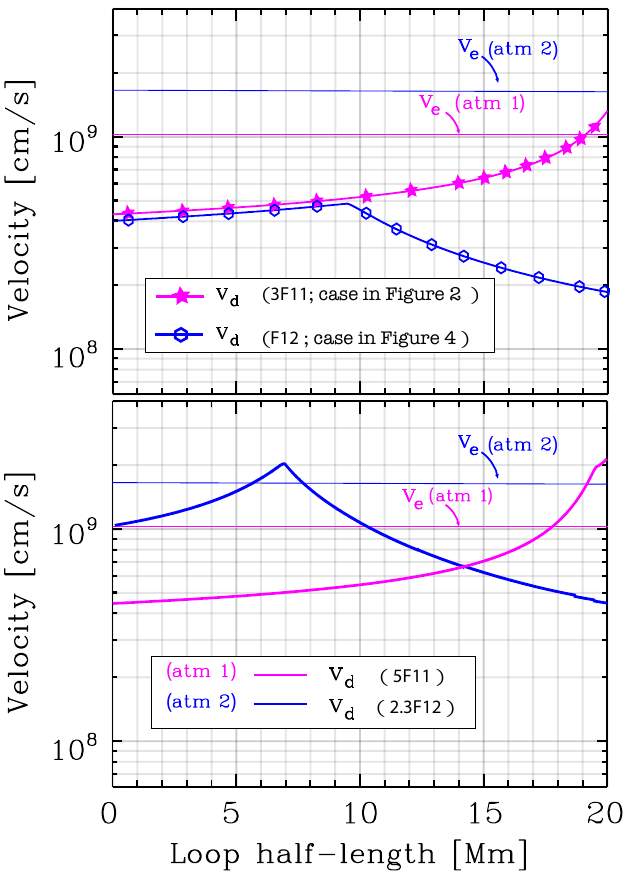}
   \caption{Comparison of the RC drift velocity and thermal speed. Upper panel: Star pink and polygon blue curves are the drift velocities for the cases in in Figures~\ref{fig:example_with_without}, ~\ref{fig:thermal}, respectively. Solid pink and blue lines are the thermal speeds in atmospheres 1 and 2, respectively. Lower panel: same as upper panel where the injected beams in atmospheres 1 and 2 satisfy the condition $n_b=0.3 n_e$ and a low energy cutoff of 20 keV. 3F11, F12, 5F11 and 2.3F12 refer to the injected energy flux density in erg cm$^{-2}$ s$^{-1}$ $3\times10^{11}$, $1\times10^{12}$, $5\times10^{11}$, $2.3\times10^{12}$, respectively.}
  \label{fig:velocity}
\end{figure}

Classical thresholds for current-driven instabilities are calculated for a current propagating in a plasma which is a simpler scenario compared with the co-evolution of the heating and the beam/RC system. \citep[e.g., Figure 1 in][]{1985ApJ...293..584H}. Another complication is that the ion temperature is also unknown and some instabilities are sensitive to the electron to ion temperature ratio, like the electrostatic ion acoustic (IA) instability \citep{1981ApJ...245..721D,1988SoPh..115..289C,1978SoPh...58..139H}, while the Buneman and electrostatic ion cyclotron (EIC) are insensitive to the temperature ratio, \citep[e.g.,][]{1977RvGSP..15..113P,1978A&A....68..145N,2002ASSL..279.....B}.

For simplicity, we consider the lowest threshold for a current-driven instability to occur if the electron and ion temperatures are comparable, $T_e\sim T_i$, which is about the electron thermal speed $v_e$, and note that the threshold is lower as the temperature ratio $T_e/T_i$ increases. It is lower by an order of magnitude if $T_e\sim5 T_i$ \citep[e.g.,][]{1985ApJ...293..584H,2002ASSL..279.....B}. Note that the thermal speed in Holman is lower than the thermal speed in this paper by a factor $\sqrt{2}$.

We compare the drift velocities to the electron thermal speed calculated in the cases discussed in previous sections. Remember that all cases in this paper satisfy the condition that the injected electron beam density, $n_b$, is at most $30\%$ of the background density in the upper corona, $n_e$.

The upper panel of Figure~\ref{fig:velocity} shows the drift speed for the beam/atmosphere cases in Figures~\ref{fig:example_with_without} and ~\ref{fig:thermal}, and the corresponding electron thermal speeds in Atmospheres 1 and 2 (see Table~\ref{tab:atm}). This shows that the drift speed is lower than the thermal speed in the second case everywhere along the loop, but the first case shows a drift-to-thermal speed ratio higher than 1 between 19 Mm and 20 Mm. It is therefore possible that the EIC instability may arise in the lower 1 Mm of the corona.

The lower panel shows that the drift speed can be higher than the thermal speed at different positions along the loop if we increase the injected energy flux density into Atmospheres 1 and 2 to $5\times10^{11}$ erg cm$^{-2}$ s$^{-1}$ and $2.3\times10^{12}$ erg cm$^{-2}$ s$^{-1}$, respectively. These values correspond to the maximum injected flux in each atmosphere used in Figure~\ref{fig:example}, i.e., for the injected beam density $n_{b0}=0.3n_e(x=0)$.

Although not shown in the figure, the drift speeds for all beam cases injected in Atmosphere 3, 4 and 5 are lower than the electron thermal speed. 

The drift velocity of the RC in the runaway model is reduced compared to purely collisional RC models. However, this does not guarantee that the drift speed is low enough to keep the beam/RC system stable to RC driven instabilities. 

If the ratio $T_e/T_i$ is initially greater than $\sim5$, a current-driven instability is likely to arise. Interestingly, if current-driven instabilities play a role in the beam/RC dynamics, the runaway electrons will be reduced because the collisions are enhanced \citep[e.g.,][]{2005SSRv..121..237B}, thereby increasing the Dreicer field. However, for most cases, the drift speed is lower than the thermal speed if $T_e\sim T_i$. 

A time-dependent treatment might be necessary for higher injected electron flux densities. 
Consider the following dynamic scenarios. The increase in the electron heating rate due to the RC, decreases the Dreicer field and increases the fraction of runaways. In higher injected flux densities, the thermalized electrons can become a significant fraction of the background electrons. This increases the Dreicer field, and results in fewer runaways. Finally, a current-driven instability may arise, thereby heating the ions and then electrons to higher temperatures. 

It is important to understand the underlying physics for higher injected flux densities because if more runaway electrons are accelerated, the heating is reduced in the corona compared to collisional RC models, whereas the heating is increased in the unstable regime. Another important distinction is the nature of electrons returning to the acceleration region, i.e., partially or fully suprathermal in the runaway case, and likely fully thermal in the unstable case.

\section{Summary}
\label{sec:summary}
The aim of this work is to investigate the conditions for which runaway electrons affect the dynamics of the nonthermal electron beam/return-current system and how. 
Although the idea that the return current can be carried by suprathermal runaway electrons has been suggested and partially explored \citep[][]{1985A&A...142..219R,1993SoPh..145..137K}, this study presents a self-consistent treatment of the electric field spatial evolution, where the co-evolution of the electron beam and components of the co-spatial RC, i.e., the thermal drift current, and the runaway current, are considered. We have explored the effects different solar loop temperature and density stratifications and injected beam distributions have on the resulting heating rate and on the number of suprathermal runaway electrons returning to the top of the loop, where acceleration is assumed to take place. Understanding return currents is necessary for an accurate inference of the energy deposition by the electron beam, and helps us constrain the accelerated beam distribution as well as the environment in which it propagates.

Our main results are:
\begin{enumerate}
\item Suprathermal electrons return to the acceleration region; they can be tens of percent of the RC flux and equivalently, the accelerated beam flux. Remember that the direct beam flux is equal to the total return current flux because of the balance between direct beam and return current densities $-J_{b}=J_{RC}=J_{d}+J_{run}$.
\item Runaway electrons can gain energies of $10-35$\,keV at the looptop. This study shows that the acceleration region is resupplied by suprathermal runaway electrons and that these electrons can gain energies up to a few tens of keV. These runaway electrons are most easily accelerated in lower temperature plasmas with $T\lesssim10$\,MK, before the loop is heated to temperatures observed in X-rays during the impulsive phase of large flares, with $T>10$\,MK. The energy gain by supra-thermal runaway electrons can be higher that 35\,keV for different choices of atmospheres, for example, a lower temperature than 3.5 MK and higher density than $4.5\times10^{9}$ cm$^{-3}$ of Atmosphere 1.
\item Neglecting runaways overestimates the heating by up to an order of magnitude compared to models of the return current in which the runaways are included. However, return current losses are rarely considered in the interpretation of X-ray spectra and the energy budget of the flare. The resulting heating rate can be significantly underestimated if only Coulomb collisional losses are assumed. 

\end{enumerate}

Secondary results:
\begin{itemize}
\item Return current heating and runaway electrons can dominate in cooler coronal loops $T<10$\, MK. The lower the background plasma temperature, the higher the total potential drop and the fraction of runaway electrons at the looptop.

\item The energy lost by the nonthermal beam is lower in the runaway model compared to a collisional (Ohmic) model of the return current, because the RC electric field is reduced by runaway electrons. In addition, the energy lost by the beam serves to both heat the ambient plasma \emph{and} accelerate a fraction into runaway electrons.

\item The RC electric field can be less than $0.1\mathscr{E}_D$ everywhere along the coronal loop and still result in a significant runaway flux at the looptop. This was shown in the Atmosphere~1 case presented in Figure~\ref{fig:example} in which for an injected flux of $2.3\times10^{18}$ cm$^{-2}$ s$^{-1}$, 20$\%$ of the RC flux at the looptop is carried by runaways, with an average energy of 10 keV. 

\item The higher the injected flux, the higher the runaways, but that also depends on the loop's electron temperature and density. For example, Atmosphere~1 and the atmosphere in Section~\ref{sec:pitch-angle} have about the same apex temperature, but different densities. A beam with $F_{e0}=6.25\times10^{18}$ electrons\,cm$^{-2}$\,s$^{-1}$ injected in the loop with varying $T$ and $n_e$ results in a higher RC total potential drop and runaway fraction. This is because the temperature decreases along the loop \emph{and} the density increases, both of which increase the number of runaways. The RC electric field is lower than 0.15$\mathscr{E}_D$, the total potential drop is 33\,kV, and the fraction of runaways is 64$\%$ at the looptop. As discussed in section~\ref{sec:pitch-angle}, both beam electrons and runaway electrons affect the number of suprathermal electrons returning to the acceleration region.

\item The runaway current acts as a self-reducing mechanism.  This can be inferred from the slower increase of the potential drop in Figure~\ref{fig:example}. Contrary to the case where runaways are neglected, the return current potential drop is reduced by a higher rate for increased injected electron beam flux density. 

\item The height along the loop of maximum energy deposition is a function of the low-energy cutoff. The higher the low-energy cutoff, the farther from the looptop is the maximum of energy deposition. That is because lower energy electrons can be thermalized along the loop.

\item The higher the loop electron density, the higher the injected beam flux is possible before moving into the super-Dreicer regime. A higher density also increases CCs in the corona which contribute to the local heating and decrease the distance of thermalization from the looptop. This also increases the X-ray emission in the corona \citep[cf.][for a review of how different mechanisms affect HXR spectra]{2011SSRv..159..107H}.

\item  Since the acceleration of runaway electrons increases non-linearly with cooler loops, higher runaways are qualitatively consistent with the beginning of the impulsive phase before the coronal loop is heated by chromospheric evaporation, or if the electron propagation has moved to a cooler loop during the impulsive phase. %\jcacomment{I know ablation is technically a better word, but nobody uses that. Chromospheric evaporation is widely used in the literature.}, 
\item If the injected electron beam density is lower than $30\%$ of the background electron density of Atmospheres 3, 4, and 5, and if the electron and ion temperatures are comparable, then the beam/RC system is likely stable to current-driven instabilities. For Atmospheres 1 and 2, the beam density must be less than $10\%$ of the background electron density for the drift velocity to be lower than the thermal speed.

\item If the electron-to-ion temperature is initially higher than 5, the reduction of the drift speed due to the presence of runaway electrons is not sufficient to keep the beam stable to current-driven instabilities for most cases presented in this paper.

\end{itemize}

Future theoretical work will include the effect of energy diffusion and pitch-angle scattering by extending the model to 2D in velocity space where the evolution of the electrons pitch-angle is taken into account in describing the beam/RC dynamics.

\section*{Acknowledgements} Meriem Alaoui is grateful to T\"unde F\"ul\"op, Ola Embreus and Mathias Hoppe for sharing their runaway codes and fruitful discussions which allowed a greater insight into the physics of runaways. We thank Arnold Benz and Brian Dennis for useful discussions. MA and JCA were supported by The Heliophysics Innovation Fund and NASA's Heliophysics Supporting Research grant 80NSSC19K0854. Numerical data is available upon request.

\end{document}